\documentclass[twocolumn,aps,showpacs,groupedaddress]{revtex4-1}
\usepackage[latin9]{inputenc}
\usepackage{amsmath}
\usepackage{amssymb}
\usepackage{graphicx}
\usepackage{bm}
\usepackage{dcolumn}
\usepackage{amsfonts}
\usepackage{color}

\makeatother

\begin{document}

\title{Topological invariant for two-dimensional open systems}

\author{Jun-Hui Zheng}
\author{Walter Hofstetter}

\affiliation{Institut f{\"u}r Theoretische Physik, Goethe-Universit{\"a}t,
60438 Frankfurt/Main, Germany.}

\date{\today}

\begin{abstract}
We study the topology of two-dimensional open systems in terms of the Green's function.  The Ishikawa-Matsuyama formula for the integer topological invariant is applied in open systems, which indicates the number difference of gapless edge bands arising from  the poles and zeros of the Green's function. Meanwhile, we define another topological invariant via the single-particle density matrix, which works for general gapped systems and is equivalent to the former
for the case of weak coupling to an environment. We also discuss two applications.  For time-reversal-invariant insulators, the ${Z}_{2}$ index can be expressed by the
invariant of each spin subsystem.  As a second application, we consider the proximity effect when an ordinary insulator is coupled to a topological insulator. 
\end{abstract}
\maketitle

\section{Introduction}

The study of the quantum Hall effect has led to the new classification paradigm of quantum phases based on topological properties\,\cite{Thouless1982prl,Wen1990prb,Hasan2010rmp,Qi2011rmp}.
The corresponding topological invariants have been constructed for various systems with interaction\,\cite{Thouless1982prl,Ishkawa1987npb,Wang2010prl,Gurarie2011prb,Niu1985prb},
with disorder\,\cite{Niu1985prb}, and with time-reversal symmetry\,\cite{Kane2005prl,Sheng2006prl,Lee2008prl}.
For the noninteracting case, the topological index is given by the integral of the Berry curvature determined by single-particle wave functions of the system\,\cite{Thouless1982prl,Wang2010prl}. For the interacting
case, it is more convenient to use the expression in terms of the single-particle Green's function\,\cite{Ishkawa1987npb} instead of the many-body wave function\,\cite{Niu1985prb}. In combination with (dynamical) mean field theory\,\cite{Georges1996rmp}, it is feasible to obtain the topological index by calculating multiple integrals of the interacting Green's function\,\cite{Wang2012epl}.
Meanwhile, two new developments have allowed us to simplify the calculation: (1) the topological Hamiltonian method, which captures topological properties of the original interacting Hamiltonian\,\cite{Wang2012prx}, and (2) the quasiparticle Berry curvature method, where the Berry curvature is determined by the quasiparticle states if these states have a long lifetime\,\cite{Shindou2008prb}.

However, all of these methods are valid only for
closed quantum systems. A realistic system, inevitably, is coupled to an environment. Especially for the noninteracting case, the ground state of an open system is expected to be a reduced density matrix (a mixed state) rather than a pure state. An extension of the Berry phase to mixed states in one-dimensional systems \cite{Uhlmann1986,Nieuwenburg2014} and related observables (e.g., Thouless pumping) \cite{Linzner2016prb,Bardyn2017} has been proposed. For two-dimensional (2D) systems, there are several different versions of topological invariants for a general density matrix\,\cite{Diehl2011np,Rivas2013prb,Bardyn2013njp,Huang2014prl,Budich2015prb,Grusdt2017}, and only one of them corresponds to the $U(1)$ holonomy\,\cite{Budich2015prb}. To include interaction effects, like for closed systems, the best option is to express the topological invariant in terms of the Green's function. 

For a general system $A$ immersed in an environment $E$, the effective partition function can be obtained by integrating out the degree of freedom of $E$. The resulting effective action retains the $U(1)$ symmetry \footnote{ The effective action is invariant under the $U(1)$ transformation $c_A \rightarrow e^{i \theta} c_A$, where $c_A$ is the Grassmann field. For the case of weak coupling to the environment, the relaxation time of density fluctuations is supposed to be short, so that the total particle number of subsystem $A$ is approximately conserved. On the other hand, in the thermodynamic limit, the fluctuations of the total particle number in system $A$ are also suppressed;, thus, the $U(1)$ symmetry, at a certain timescale, is obtained.
}.  
Accordingly, the first Chern number of the system $A$ can be defined by the Ishikawa-Matsuyama formula,
\begin{equation}
\text{Ch}_{A}=\frac{\varepsilon^{\mu\nu\rho}}{24\pi^{2}}\int d^{3}k\text{Tr}[{G}_{A}\partial_{\mu}{G}_{A}^{-1}{G}_{A}\partial_{\nu}{G}_{A}^{-1}{G}_{A}\partial_{\rho}{G}_{A}^{-1}],\label{Chern}
\end{equation}
where $\mu$, $\nu$, and $\rho$ run through $k_{0}$, $k_{1}$, and $k_{2}$, with $k_{0}=i\omega$ being the imaginary frequency, and the Matsubara Green's function is
$G_{A}^{\sigma\sigma^{\prime}}(\mathbf{k},\tau-\tau^{\prime}) = -\langle T_{\tau}\hat{c}_{A,\mathbf{k}\sigma}(\tau)\hat{c}_{A,\mathbf{k}\sigma^{\prime}}^{\dagger}(\tau^{\prime})\rangle$.
Here, $\sigma$ and $\sigma'$ represent the internal degrees of freedom of  $A$. The translational symmetry is assumed for  both  $A$ and  $E$. The index (\ref{Chern}) is a well-defined topological invariant\,\cite{Qi2008prb}  iff there are neither poles nor zeros for $G_{A}(\mathbf{k},i\omega)$ for all $\mathbf{k}$ and imaginary  frequency $i\omega$.

We develop both the topological Hamiltonian method and the Berry curvature method to evaluate Eq.\,(\ref{Chern}). The topological Hamiltonian gives an effective single-particle description of the open system. On the other hand, the Berry curvature method shows that  for  an open system,
besides the bands determined by the poles of the
Green's function (energy spectrum), the ``bands'' $\omega(\mathbf{k})$ from the zeros (denoted as {\it blind} bands here), i.e., $\text{det}G_{A}(\mathbf{k},\omega(\mathbf{k})) = 0$, appear and contribute to the topological properties. The value of  $\text{Ch}_{A}$ indicates the number difference of gapless edge modes and gapless edge blind bands. 

\begin{figure}
\centering \includegraphics[width=3.1 in]{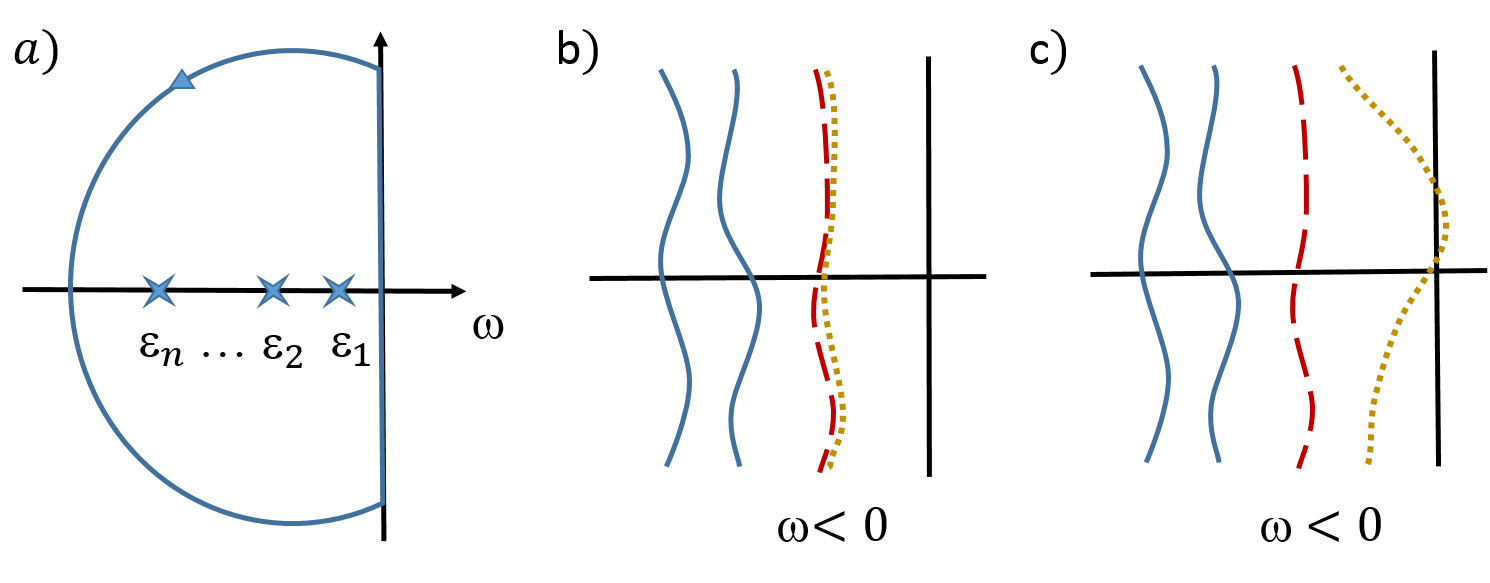} \caption{(a) The path for the contour integral in the $\omega$ plane. The band structures for (b) the weak-coupling case and (c) for the strong-coupling case.
The solid blue lines refer to the poles with $j\in\mathcal{O}_A$, the dashed red lines refer to the new bands ($j\notin\mathcal{O}_A$) due to the coupling to the environment, and the  dotted yellow lines are blind bands.}
\label{fig1} 
\end{figure}

When the coupling to the environment becomes larger, the blind bands may cross the Fermi surface [Fig.\ref{fig1}(c)],  i.e., $\text{det}G_{A}(\mathbf{k}, \omega=0) =0$
for some $\mathbf{k}$, and thus, $\text{Ch}_{A}$
becomes ill defined even though the energy spectrum is still gapped. For a general gapped system, we propose a topological invariant ${I}_A$  based on the single-particle density matrix $\rho_{A}(\mathbf{k})$ with $\rho_{A}^{\sigma\sigma'}=\langle c_{A,\mathbf{k}\sigma'}^{\dagger}c_{A,\mathbf{k}\sigma}\rangle$, which is equivalent to $\text{Ch}_{A}$ when the system
is weakly coupled to the environment.  For a noninteracting  system, there is a  one-to-one correspondence between the density matrix and the entanglement Hamiltonian\,\cite{Peschel2002,Fidkowski2010prl}. Unlike the previous discussions of the relation between the entanglement spectrum (entropy) and topological properties of the full system ($A+E$)\,\cite{Kitaev2006prl,Hsieh2014prl,Chandran2014prl}, we focus on the physical implication of the eigenstates of the density matrix for $A$. The index ${I}_A$ [see Eq.\,\eqref{chern_a} below] is determined by the Berry curvature of the dominant eigenstates of the density matrix $\rho_{A}(\mathbf{k})$, as proposed in Ref.\,\cite{Budich2015prb} for noninteracting systems,  giving the number of gapless edge modes for $\rho_A$.

This paper is organized as follows. In Sec. \ref{thm}, we show that the topological Hamiltonian is applicable for the case when $G_{A}(\mathbf{k},i\omega)$ has neither poles nor zeros. In Sec. \ref{bcm}, we express the Chern number (\ref{Chern}) through the Berry curvature of the bands of the energy spectrum and the blind bands.  In Sec. \ref{dmd}, a density-matrix description of topological invariance is developed. In Sec. \ref{app}, we discuss the applications of our theory.

\section{Topological Hamiltonian method} \label{thm}

 We suppose that $G_{A}(\mathbf{k},i\omega)$ in Eq.\,\eqref{Chern}  has neither poles nor zeros and try to find the corresponding topological Hamiltonian through smooth deformation\,\cite{Wang2012prx}. The deformation path is set to be $P_{\lambda}$: ${G}_{A}(\mathbf{k},i\omega,\lambda)=(1-\lambda){G}_{A}(\mathbf{k},i\omega)+\lambda[i\omega+{G}_{A}^{-1}(\mathbf{k},i\omega=0)]^{-1}$ for $\lambda\in[0,1]$. For $i\omega=0$, we have ${G}_{A}(\mathbf{k},i\omega,\lambda)={G}_{A}(\mathbf{k},i\omega)$, which is presumed to have neither poles nor zeros.  For $i\omega\neq0$, like in Ref.\,\cite{Wang2012prx},  using the
Lehmann representation of $G_A$ in the zero-temperature limit, Eq.\,\eqref{green},  it is easy to check that $\text{sgn}[\text{Im}\langle a|{G}_{A}(\mathbf{k},i\omega,\lambda)|a\rangle]=-\text{sgn}(\omega)$ for any vector $|a\rangle$ in the subspace $A$, which implies that (the imaginary part of)  every eigenvalue of ${G}_{A}(\mathbf{k},i\omega,\lambda)$ is nonzero.  Note that ${G}_{A}(\mathbf{k},i\omega,\lambda)$ also cannot diverge for a well-defined ${G}_{A}(\mathbf{k},i\omega)$. Therefore, during the deformation, $\text{Ch}_A$ remains well defined, and thus, the deformation $\mathcal{P}_{\lambda}$
is smooth. This means that the single-particle topological Hamiltonian $H_{A}^{\text{topo}}(\mathbf{k})\equiv-G_{A}^{-1}(\mathbf{k},i\omega=0)$, whose Green's function is ${G}_{A}(\mathbf{k},i\omega,\lambda=1)$, has the same topological properties as $A$.

On the other hand, the Green's function for the full system ($A+E$) can be written as a block matrix,
\begin{equation}
G_{F}(\mathbf{k},\tau-\tau^{\prime})=\begin{pmatrix}G_{A} & G_{AE}\\
G_{EA} & G_{E}
\end{pmatrix},\label{eq:green_ful}
\end{equation}
where $G_{A}^{\sigma\sigma^{\prime}}(\mathbf{k},\tau-\tau^{\prime})=-\langle T_{\tau}\hat{c}_{A,\mathbf{k}\sigma}(\tau)\hat{c}_{A,\mathbf{k}\sigma^{\prime}}^{\dagger}(\tau^{\prime})\rangle$,  $G_{AE}^{\sigma\eta} = - \langle T_{\tau}\hat{c}_{A,\mathbf{k}\sigma}(\tau)\hat{c}_{E,\mathbf{k}\eta}^{\dagger}(\tau^{\prime})\rangle$,  $G_{EA}^{\eta\sigma} = -\langle T_{\tau}\hat{c}_{E,\mathbf{k}\eta}(\tau)\hat{c}_{A,\mathbf{k}\sigma}^{\dagger}(\tau^{\prime})\rangle$, and
 $G_{E}^{\eta\eta^{\prime}} = -\langle T_{\tau}\hat{c}_{E,\mathbf{k}\eta}(\tau)
 \hat{c}_{E,\mathbf{k}\eta^{\prime}}^{\dagger}(\tau^{\prime})\rangle$. The indices $\eta$ and $\eta^{\prime}$ refer to the internal degrees of freedom for the environment. Introducing the projection operator $P_{A}$ for the $A$ subspace, we have $G_{A}=P_{A}G_{F}P_{A}$. Immediately, we obtain the relation $H_{A}^{\text{topo}}=[P_{A}(H_{F}^{\text{topo}})^{-1}P_{A}]^{-1}$, where $H_{F}^{\text{topo}}(\mathbf{k})=-G_{F}^{-1}(\mathbf{k},i\omega=0)$ is the topological Hamiltonian of the full system.  In particular, for a noninteracting system, $H_{F}^{0}$, the topological property for the system $A$ is determined by $[P_{A}(H_{F}^{0})^{-1}P_{A}]^{-1}$ instead of $P_{A}H_{F}^{0}P_{A}$.  The former contains the information on the coupling between $A$ and $E$.

 \section {Berry curvature method} \label{bcm} 

 For the integral (\ref{Chern}), considering
the contour integral in Fig.\,\ref{fig1}(a) for $\omega$ and using  the
residue theorem, the result depends only on the behavior of
the Green's function at its poles and zeros (see Appendix A).
Generally, the integral (\ref{Chern}) gives
\begin{eqnarray}
\text{Ch}_{A} & = & \sum_{j=1}^{n}\frac{\varepsilon^{\alpha\beta}}{2\pi i}\int d^{2}k\langle\partial_{\alpha}\psi_{j}^{A}(\mathbf{k})|\partial_{\beta}\psi_{j}^{A}(\mathbf{k})\rangle\nonumber \\
 &  & -\sum_{j=1}^{n-m}\frac{\varepsilon^{\alpha\beta}}{2\pi i}\int d^{2}k\langle\partial_{\alpha}\phi_{j}^{A}(\mathbf{k})|\partial_{\beta}\phi_{j}^{A}(\mathbf{k})\rangle.\label{berry_a}
\end{eqnarray}
where $|\psi_{j}^{A}(\mathbf{k})\rangle$ and $|\phi_{j}^{A}(\mathbf{k})\rangle$ are the eigenvectors of the Green's function $G_{A}(\mathbf{k},\omega)$ with divergent eigenvalues (i.e., poles of $G_{A}$) and zero eigenvalues (i.e., zeros  of $G_{A}$) for $\omega<0$, respectively. Here, we have supposed that given a momentum the number of the poles is $n$ and that of the zeros is $n-m$ for $\omega <0$. Clearly, the value of  $\text{Ch}_{A}$ indicates the difference between the number of gapless edge modes and gapless edge blind bands.

Now, we compare Eq.\,\eqref{berry_a} with the first Chern number of a closed system: the full system.  For the full system, the formula of the first Chern number is similar to Eq.\,\eqref{Chern}, but with $G_A$ replaced by $G_F$.  

 For the full system, the zeros of $G_{F}(\mathbf{k},\omega)$ may appear only for strong interactions with some special symmetries\,\cite{Gurarie2011prb}, and then the Berry curvature method will give a result similar to that of Eq.\,\eqref{berry_a}.  In most cases, $G_F(\mathbf{k},\omega)$ has no zeros. We suppose that there are, in total, $N$ bands for the full system, and the first $n$ bands are filled. Then the first Chern number of the full system is described by the $n$  filled (quasiparticle) states,
\begin{equation}
\text{Ch}_{F}=\sum_{j=1}^{n}\frac{\varepsilon^{\alpha\beta}}{2\pi i}\int d^{2}k\langle\partial_{\alpha}\psi_{j}(\mathbf{k})|\partial_{\beta}\psi_{j}(\mathbf{k})\rangle,\label{eq:berry_ful}
\end{equation}
where $\alpha$ and $\beta$ run through $k_{1}$ and $k_{2}$ (see Appendix A)\,\cite{Shindou2008prb,Wong2013prl}.  Note that the Green's function is $G_{F}(\mathbf{k},\omega)={1}/{[\omega-H_{F}(\mathbf{k},\omega)]}$, where $H_{F}(\mathbf{k},\omega)=H_{F}^{0}(\mathbf{k})+\Sigma(\mathbf{k},\omega)$,  with $\Sigma(\mathbf{k},\omega)$ being the self-energy. The poles $\varepsilon_{j}(\mathbf{k})$
are determined by $\text{det}[H_{F}(\mathbf{k},\varepsilon_{j}(\mathbf{k}))-\varepsilon_{j}(\mathbf{k})]=0$. The  state $|\psi_{j}(\mathbf{k})\rangle$ in Eq.\,\eqref{eq:berry_ful} is the eigenstate
of $H_{F}(\mathbf{k},\omega=\varepsilon_{j}(\mathbf{k}))$ with the eigenvalue $\varepsilon_{j}(\mathbf{k})<0$ (numerically, $\varepsilon_{j}(\mathbf{k})$ is the value of $\omega$  where $\text{det}[H_{F}(\mathbf{k},\omega)-\omega]$ changes sign).  Here, we have supposed that each band has a long lifetime to use the concept of quasiparticle, which breaks down for strong-interaction cases.

For the noninteracting case, the states in the set $\mathcal{R} \equiv \{|\psi_{j}(\mathbf{k})\rangle\}_{ j=1,\cdots,n}$ are orthogonal  to each other, and Eq.\,\eqref{eq:berry_ful} is invariant under  a $U(n)$
transformation for $|\psi_{j}\rangle$ in $\mathcal{R}$  (see Appendix B).  For the interacting case, these states are not orthogonal anymore, but with the assumption that they are still linearly independent (which is expected for weak and moderate interactions) and using Gram-Schmidt orthogonalization,  we prove that any orthogonal
basis of  the set $\mathcal{R} $ gives exactly the same Chern number (\ref{eq:berry_ful}) in Appendix. As a result, the Chern number is related only to the space spanned by the set $\mathcal{R} $ for both cases. This finding allows us to simplify the problem. 

Choosing the contour integral for the Green's function $G_F$ as shown in Fig.\,\ref{fig1}(a), we get the following relation: 
\begin{equation} \small
\sum_{j=1}^{n}\frac{\langle\mathbf{k},l|\psi_{j}(\mathbf{k})\rangle\langle\psi_{j}(\mathbf{k})|\mathbf{k},l'\rangle}{1-\partial_{\omega}E_{j}(\mathbf{k},\omega)}{\Big{|}}_{\omega=\varepsilon_j(\mathbf{k})}=\frac{1}{2\pi i}\oint dze^{z\tau}G_{F}^{l,l'}(\mathbf{k},z),\label{eq:space_ful}
\end{equation}
where $\tau=0^{+}$ and $l$ includes all of the internal degrees of freedom ($\sigma$ and $\eta$). The energy $E_{j}(\mathbf{k},\omega)$ in the denominator is the $j$th eigenvalue of $H_{F}(\mathbf{k},\omega)$, which satisfies $E_{j}(\mathbf{k},\varepsilon_{j}(\mathbf{k}))=\varepsilon_{j}(\mathbf{k}) $.  In addition,
$\frac{1}{2\pi i}\oint d z e^{z\tau}G_{F}^{l,l'}(\mathbf{k},z) =\frac{1}{2\pi}\int d\omega e^{i\omega\tau}G_{F}^{l,l'}(\mathbf{k}, i\omega)
=\langle c_{\mathbf{k}l'}^{\dagger}c_{\mathbf{k}l}\rangle$. Thus, from Eq.\,\eqref{eq:space_ful}, for the single-particle density matrix  $\rho_{F}(\mathbf{k})$ with $\rho_{F}^{ll'}(\mathbf{k})=\langle c_{\mathbf{k}l'}^{\dagger}c_{\mathbf{k}l}\rangle$, the eigenstate set $\{|\overline{\psi}_j(\mathbf{k})\rangle\}_{j=1,\cdots,n}$ of $\rho_F(\mathbf{k})$ with nonzero eigenvalues,  spans exactly the same space as the set $\mathcal{R}$. Consequently, the Chern number (\ref{eq:berry_ful}) equals the sum of the integral of Berry curvature  of  all states  $|\overline{\psi}_j(\mathbf{k})\rangle$. This provides a method for evaluating the topological index beyond  the topological Hamiltonian method.

  For an open system $A$,  in general, $\text{{det}}G_{A}(\mathbf{k},\omega)=0$ at some frequencies.  To see the emergence of these blind bands, let us for simplicity focus on the noninteracting case. We still suppose that there are in total $N$ bands for the full system, and the first $n$ bands are filled. For the case that the system $A$ decouples from the environment $E$, we now prove that no blind bands exist. Let the set $\mathcal{R}_A \equiv \{|\psi_{j}(\mathbf{k})\rangle\}_{j \in \mathcal{O}_A}$, with $\mathcal{O}_A \equiv \{1,\cdots,m, n+1,\cdots,n+m'\}$ (i.e., $m$ filled bands and $m^\prime$ unfilled bands) belong to the system $A$ and the other bands $\mathcal{R}_E \equiv \{|\psi_{j}(\mathbf{k})\rangle\}_{j \in \mathcal{O}_E}$, with $\mathcal{O}_E \equiv \{m+1,\cdots,n, n+m'+1,\cdots,N\}$ (i.e., $n-m$ filled bands and $N-n-m^\prime$ unfilled bands)  belong to the environment. Set $|\varphi_{j}(\mathbf{k})\rangle \equiv P_{A}|\psi_{j}(\mathbf{k})\rangle$, then
 $|\varphi_{j}(\mathbf{k})\rangle =0$ for $j \in \mathcal{O}_E$, and the set  $ \{|\varphi_{j}(\mathbf{k})\rangle\}_{j \in \mathcal{O}_A}$ forms a complete orthogonal set in the subspace $A$.  So the Green's function becomes  $G_{A}(\mathbf{k}, \omega)= \sum_{j\in \mathcal{O}_A}\frac{|\varphi_{j} (\mathbf{k})\rangle\langle\varphi_{j} (\mathbf{k})|}{\omega-\varepsilon_{j}(\mathbf{k})}$. If $G_{A}(\mathbf{k},\omega)|u\rangle =0$ for a nonzero $|u\rangle$, this  implies $\{|\varphi_{j}(\mathbf{k})\rangle\}_{j \in \mathcal{O}_A}$  are linearly dependent, which contradicts the orthogonality. Therefore, the Green's function $G_A$ has no zero eigenvalues.

When the system is weakly coupled to the environment, we have $G_{A}(\mathbf{k},\omega)=\sum_{j=1}^{N}\frac{|\varphi_{j}(\mathbf{k})\rangle\langle\varphi_{j}(\mathbf{k})|}{\omega-\varepsilon_{j}(\mathbf{k})}$,
and $|\varphi_{j}(\mathbf{k})\rangle$ now can be nonzero even for $j \in \mathcal{O}_E$.  Hence, the new poles $\varepsilon_{j}(\mathbf{k})$ from $j \in \mathcal{O}_E$ emerge,  accompanied by the appearance of the zeros of $G_A$\,\cite{Gurarie2011prb}. The latter cancels the effects from the new bands [see eq.\,\eqref{berry_a}], which is consistent with the robustness of the topological index.  Intuitively, the set $\{|\varphi_{j}(\mathbf{k})\rangle\}_{j=1,\cdots,N}$ is overcomplete, and  it allows $|u\rangle$ to be an eigenstate of $G_{A}(\mathbf{k},\omega)$ with zero eigenvalue. Note that for the new poles $\varepsilon_{j}$ with $j \in \mathcal{O}_E$, one has
$\mathcal{V}_{j}\equiv\langle\varphi_{j}(\mathbf{k})|\varphi_{j}(\mathbf{k})\rangle \ll 1$ for the weak-coupling case.  Thus, for $\omega$ away from $\varepsilon_{j}$ ($j \in \mathcal{O}_E$), the weak coupling corrects the Green's function slightly and cannot contribute zeros. However, if $\omega-\varepsilon_{j}\sim \mathcal{V}_{j}$ for $j \in \mathcal{O}_E$,  the correction becomes significant. As a result, for the weak-coupling case, the blind bands will always be close to those new bands [Fig.\ref{fig1}(b)], and actually, they always appear in pairs\,\cite{Gurarie2011prb}. Consequently, below the Fermi surface, the number of blind bands and that of the new bands are the same. The number of effective bands occupied in the system $A$ is $m \equiv N_{pole}-N_{blind}$, where $ N_{pole}$ and $N_{blind}$ are the number of energy bands and blind bands below the Fermi surface, respectively. 

\section{ Density-matrix description} \label{dmd}

 For weak coupling and the noninteracting case, the blind
bands are always close to the new poles. This property allows us to smoothly tune $\varepsilon_{j}(\mathbf{k})$ for a well-defined $G_A(\mathbf{k},\omega)$, avoiding bands or blind bands crossing the zero point,
so that the first $n$ poles [$\varepsilon_{j}(\mathbf{k})<0$] are moved together to $\varepsilon_{G}(\mathbf{k})<0$,  and the others are moved to $\varepsilon_{E}(\mathbf{k})>0$, without changing the  topological index (\ref{Chern}). The final Green's function becomes $\underline{G}_{A}(\mathbf{k},\omega)=\frac{\rho_A (\mathbf{k})}{\omega-\varepsilon_{G}(\mathbf{k})}+ \frac{\overline{\rho}_A(\mathbf{k})}{\omega-\varepsilon_{E}(\mathbf{k})}$,
where $\rho_{A}(\mathbf{k})=\sum_{j=1}^{n}|\varphi_{j}(\mathbf{k})\rangle\langle\varphi_{j}(\mathbf{k})|$ and  $\overline{\rho}_{A}(\mathbf{k})=\sum_{j=n+1}^{N}|\varphi_{j}(\mathbf{k})\rangle\langle\varphi_{j}(\mathbf{k})|$.  By using the completeness $\sum_{j=1}^{N}|\psi_{j}(\mathbf{k})\rangle\langle\psi_{j}(\mathbf{k})|=\mathbf{I}$, we find $\overline{\rho}_A =\mathbf{I}_A - \rho_A$,  where $\mathbf{I}_A$ is the unit matrix in the subspace $A$. In addition, we have $\rho_{A}^{\sigma\sigma'}=\frac{1}{2\pi i}\oint dze^{z\tau}G_{A}^{\sigma\sigma'}(\mathbf{k},z)=\langle c_{A,\mathbf{k}\sigma'}^{\dagger}c_{A,\mathbf{k}\sigma}\rangle$.

The topological Hamiltonian for the final Green's function  $\underline{G}_{A}(\mathbf{k},i\omega)$ is  $\underline{H}_{A}^{\text{topo}} = [\varepsilon_{E}^{-1} \mathbf{I}_A - x \rho_{A}]^{-1}$, where $x = \varepsilon_{E}^{-1}-\varepsilon_{G}^{-1} >0$. Consequently, the eigenstates of $\underline{H}_{A}^{\text{topo}}$ with negative eigenvalues are exactly the eigenstates of $\rho_{A}$ with  the largest $m$ eigenvalues.
For the interacting case, using the Lehmann representation of the Green's function $G_A$ and by properly tuning the position of the many-body eigenenergy, a similar discussion can be applied and the same conclusion can be obtained (see Appendix D). 

For strong coupling, the blind bands may cross the zero point even for a gapped system, in which case the Chern number (\ref{Chern}) becomes ill defined.  To extend the concept of the topological index to a general gapped system, we propose the following formula as a generalized
topological index:
\begin{equation}
{I}_{A}=\sum_{j=1}^{m}\frac{\varepsilon^{\alpha\beta}}{2\pi i}\int d^{2}k\langle\partial_{\alpha}\overline{\psi}_{j}(\mathbf{k})|\partial_{\beta}\overline{\psi}_{j}(\mathbf{k})\rangle,\label{chern_a}
\end{equation}
where $|\overline{\psi}_{j}(\mathbf{k})\rangle$ are the eigenstates of $\rho_{A}(\mathbf{k})$ with the $m$ largest eigenvalues. From the above discussion, for the weak-coupling case, we have ${I}_{A}=\text{Ch}_{A}$. 
Moreover,  formula (\ref{chern_a})  is the same as the definition of topology for a `density matrix' in\,
\cite{Bardyn2013njp,Budich2015prb}.  For the strong-coupling case with a well-defined $\rm{Ch}_A$, these two quantities may be different, if the system in this phase region cannot be mapped to an isolated system by  smoothly changing (without the blind bands and the energy spectrum crossing the zero point) the parameters such as the coupling strength.

Without interaction and without coupling to the environment, the eigenvalues of $\rho_A$ are $1$ (with degeneracy $m$) and $0$, which gives the gapped flat-band structure. The interaction and coupling will change the single-particle distribution, and the null eigenvalues become finite. However, the $m$ majority eigenstates are still gapped from the minority eigenstates. The topological phase transition
 occurs when the $m$th largest eigenvalue of $\rho_{A}(\mathbf{k})$ becomes degenerate with the $(m+1)$th largest eigenvalue of $\rho_{A}(\mathbf{k})$ or the gap of the energy spectrum closes (which may lead to an energy band inversion for $G_A$ and thus a sudden change in the eigenstates of $\rho_A$).  In contrast, by the definition of  the Chern number (\ref{Chern}), the topological phase transition can occur when the blind bands close and reopen a gap between them without any energy band inversion\,\cite{Gurarie2011prb}.

\section{Applications}\label{app}
\subsection{ $Z_2$ topological invariant}
For a time-reversal ($\mathcal{T}$)-invariant 2D
system, the ${Z}_{2}$ topological index in the form of a Green's function is a five-dimensional integral\,\cite{Wang2010prl}. For the case with conserved spin, the ${Z}_2$ index $\nu$ can be expressed through the topological index of the decoupled spin subsystem [see Eq.\,\eqref{Chern}], i.e., $\nu= \text{Ch}_{\uparrow}~ \text{mod}~2$ \cite{Kane2005prl,Hohenadler2013jpcm,Wu2006prl,Xu2006prb},
which simplifies the calculation greatly.
Here, the Chern number $\text{Ch}_{\sigma}$ reflects
the number of edge states for each spin. The spin conservation can be violated by different types of spin-orbit coupling (SOC)\,\cite{Kane2005prl}. In the following, we point out that  for the weak SOC case, $\nu= \text{Ch}_{\uparrow}~ \text{mod}~2$,  and for a general gapped case, $\nu= {I}_{\uparrow}~ \text{mod}~2$, relating the topology of the subsystem and that of the full system \cite{Prodan2009prb,Fukui2014jpsj}. 

\begin{figure}
\includegraphics[width=3.3in]{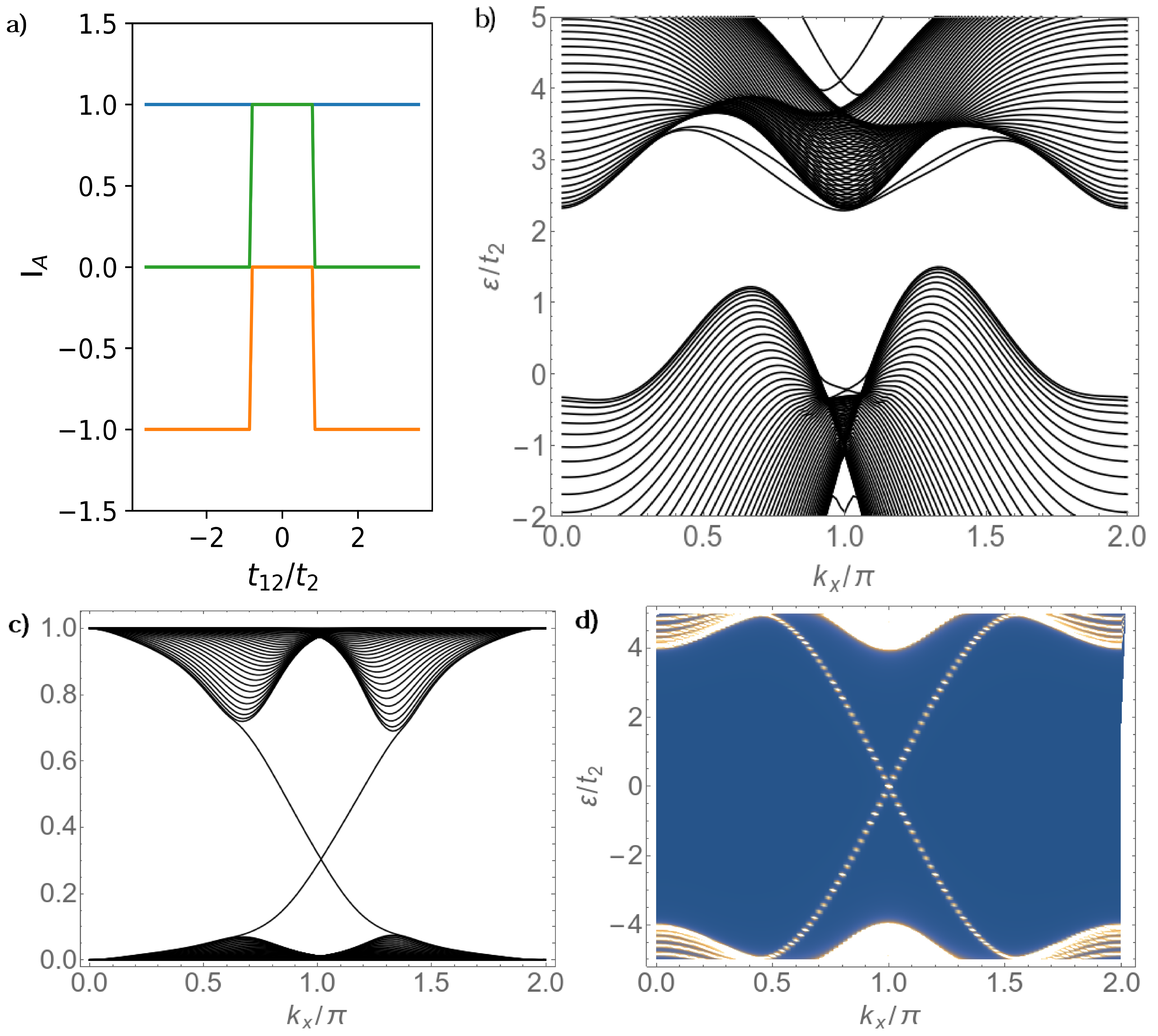}
 \caption{(Color online) (a) Topological index for the bilayer system consisting of an ordinary insulator and a Chern insulator. The green line is the Chern number for the full system. The blue line is the index ${I}_{A}$ for the
first layer, and the orange line is for the second layer.
(b) The band structure for the full system with zigzag edges.
(c) The spectrum of the density matrix for the second layer.
d) The blind bands of the Green's function for the second
layer. Here, $t_{12}=2.5t_{2}$ for (b)-(d).}
\label{topoPhase} 
\end{figure}

Let $G_{\uparrow}$ and $G_{\downarrow}$ be the Green's functions for each spin subsystem. Due to $\mathcal{T}$ symmetry, we have the following relations (see Appendix C):
$G_{\uparrow}^{*}(\mathbf{k}, i\omega)=G_{\downarrow}(-\mathbf{k},-i\omega)$,
$\rho_\uparrow^*(\mathbf{k}) = \rho_\downarrow(-\mathbf{k})$, and $G_{\uparrow}(\mathbf{k}, \omega+i\delta)=G_{\downarrow}^{T}(-\mathbf{k}, \omega+i\delta)$. The first two relations imply $\text{Ch}_{\uparrow}=-\text{Ch}_{\downarrow}$ and  ${I}_{\uparrow}=-{I}_{\downarrow}$ from Eqs.\,\eqref{Chern} and\,\eqref{chern_a}. The last relation shows a close relation between the density of state for each subsystem, which indicates that the subsystems and the full system close and reopen a gap at the same time. The topological state for the full system is changed
when the topological properties in the subsystem
are changed due to the SOC coupling to the other subsystem. Thus, the $Z_2$ index can be obtained from the index $I_A$ of the subsystem. 

In Fig.\,\ref{subsystem}
in appendix C, we plot the numerical results for the topological index ${I}_{A}$ of the subsystem of the Kane-Mele model, showing the exact relation to the ${Z}_{2}$ index, $\nu= {I}_{\uparrow}~ \text{mod}~2$.

 \subsection{ Proximity effect} 

A simple but fundamental question is what happens
when an ordinary insulator is coupled to a topological insulator. As an example, here we consider a system on a bi-layer honeycomb lattice, with each layer described by the Haldane model\,\cite{Haldane1988prl}:
$H_{\alpha} = [-t_{1} \sum_{\langle i,j\rangle}  c_{i\alpha}^{\dagger} c_{j\alpha} -t_{2}\sum_{\langle\langle i,j\rangle\rangle}e^{i\theta\phi_{\alpha}}c_{i\alpha}^{\dagger}c_{j\alpha}+ {\rm H.c.}]
 +m_{\alpha}\sum_{i}(-1)^{i}c_{i\alpha}^{\dagger}c_{i\alpha}$,
where $t_{1}$ is the nearest-neighbor tunneling, $t_{2}$ is the next-nearest-neighbor tunneling with a phase $\theta\phi_{\alpha}$ and $\theta=\pm 1$, $m_{\alpha}$ is the staggered potential, and $\alpha=1,2$ is the layer index. The parameters are $t_{1}=4t_{2}$, $\phi_{1}=\pi/2$, $m_{1}=0$, $\phi_{2}=0$, and $m_{2}=0.2t_{2}$, so that the first layer is a Chern insulator, and the second one is an ordinary insulator. 

To observe the interference effect,  we further introduce a tunneling term between layers,
$H_{12}=t_{12}\sum_{i}c_{i2}^{\dagger}c_{i1} + \rm{H.c}$.
The topological index $\text{I}_{A}$ with respect to the tunneling coefficient $t_{12}$ is shown in
Fig.\,\ref{topoPhase}a. In the whole region of $t_{12}/t_2$, the index $I_A$ for the first layer is unchanged. 
For the second layer, there are no edge modes or edge blind bands when $t_{12} =0$. For a small finite  $t_{12}$, since the full system has gapless poles on the edge, the  second layer also has a gapless edge state and simultaneously a gapless blind state appear. The Chern number $\text{Ch}_{A} = {I}_{A}$ is still zero, like as the  new bulk-boundary correspondence we claimed. The second layer subsystem obtains a nonzero topological index when $t_{12} \gtrsim 0.83 t_2$, while the full system displays a transition from a Chern insulator to a  trivial one through  band inversion.  For $t_{12}\gtrsim0.83t_2$,  the stable edge state disappears, but the gapless blind band remains, and thus, $I_A = -1$ for  the second layer.  For $t_{12}=2.5t_2$,  the spectrum of the full system with zigzag edges is shown in Fig.\,\ref{topoPhase}(b), which has no edge states. However, the spectrum of the density matrix for each layer contains a gapless edge state, which is shown in Fig.\,\ref{topoPhase}(c). For the second layer, the blind bands become gapless at the edge [Fig.\,\ref{topoPhase}(d)].

\section{ Conclusion} We have systemically discussed the topology of 2D open systems. The invariant given by the Ishikawa-Matsuyama formula reflects the number difference of gapless edge modes and gapless edge blind bands.
Moreover, we defined another topological invariant in terms of the single-particle density matrix which  is applicable  for general gapped systems and is equivalent to the former invariant for the case of weak coupling  to the environment.
Two applications were discussed.  For time-reversal invariant insulators, we explained that the relation of the 
Chern invariant for each spin subsystem and the ${Z}_{2}$ index of the full system is given by $\nu= {I}_{\uparrow}~ \text{mod}~2$, which highly simplifies the calculation for the ${Z}_{2}$ index. In addition, we consider the proximity effect when an ordinary insulator is coupled to a topological insulator, which shows an inverse topological invariant is induced by the nontrivial part \cite{Hsieh2016prl}. 
 
The examples given here are for non-interacting cases. The method can be applied to the interacting system, also at finite temperature, and the single-particle density matrix  can be  obtained, for example, by using dynamical mean-field theory.  

{ \bf Acknowledgment}:    This work was supported by the Deutsche Forschungsgemeinschaft via DFG FOR 2414 and the high-performance computing center LOEWE-CSC.

\appendix

\section{ Berry curvature method}

In this section, we derive the expression of the first Chern number in terms of Berry curvature [i.e.,  Eqs.\,\eqref{berry_a} and \eqref{eq:berry_ful} in the main text], following the method developed
in\,\cite{Shindou2008prb}. To calculate the integral
\begin{eqnarray}
\text{Ch}&=&\frac{\varepsilon^{\mu\nu\rho}}{24\pi^{2}}\int d^{3}k\text{Tr}[{G}\partial_{\mu}{G}^{-1}{G}\partial_{\nu}{G}^{-1}{G}\partial_{\rho}{G}^{-1}] \notag\\
&=&-\frac{\varepsilon^{\mu\nu\rho}}{24\pi^{2}}\int d^{3}k\text{Tr}[\partial_{\mu}{G}\partial_{\nu}{G}^{-1}{G}\partial_{\rho}{G}^{-1}],\label{s_1}
\end{eqnarray}
it is helpful to introduce a similarity transformation $U$, so that $G=UG_{d}U^{-1}$, where $G_{d}$ is diagonalized. Substituting it into Eq.\,(\ref{s_1}), we obtain 
\begin{widetext}
\begin{eqnarray}
\text{Ch} & = & -\frac{\varepsilon^{\mu\nu\rho}}{24\pi^{2}}\int d^3 k \text{Tr}[\{(\partial_{\mu}U)G_{d}U^{-1}+U(\partial_{\mu}G_{d})U^{-1}+UG_{d}(\partial_{\mu}U^{-1})\}\nonumber \\
 &  & \{(\partial_{\nu}U)G_{d}^{-1}U^{-1}+U(\partial_{\nu}G_{d}^{-1})U^{-1}+UG_{d}^{-1}(\partial_{\nu}U^{-1})\}UG_{d}U^{-1}\nonumber \\
 &  & \{(\partial_{\rho}U)G_{d}^{-1}U^{-1}+U(\partial_{\rho}G_{d}^{-1})U^{-1}+UG_{d}^{-1}(\partial_{\rho}U^{-1})\}].\label{s_2}
\end{eqnarray}
There are 27 terms in total in Eq.\,\eqref{s_2}. Twelve of the terms give total derivatives, three terms vanish due to asymmetrization, and six terms cancel with each other. The remaining terms are 
\begin{eqnarray}
\text{Ch} = \frac{\varepsilon^{\mu\nu\rho}}{4\pi^{2}}\int d^{3}k\text{\text{Tr}}[(G_{d}^{-1}\partial_{\mu}G_{d})(\partial_{\nu}U^{-1})(\partial_{\rho}U)] = \frac{\varepsilon^{\mu\nu\rho}}{4\pi^{2}}\oint dz\int d^{2}ke^{z0^{+}}\text{\text{Tr}}[(G_{d}^{-1}\partial_{\mu}G_{d})(\partial_{\nu}U^{-1})(\partial_{\rho}U)].\label{s_3}
\end{eqnarray}
Here, the contour integral in the complex plane of $\omega$   
(denoted by $z$ here) is shown in Fig.\ref{fig1}(a) in the main text. The poles and zeros of the Green's function appear only at real frequencies. For real frequency, the Green's function is Hermitian, and the elements of $U$ are $U_{\sigma j}=\langle\mathbf{k},\sigma|\psi_{j}(\mathbf{k},\omega)\rangle$, where $|\psi_{j}(\mathbf{k},\omega)\rangle$ is the $j$th eigenstate of $G$ with eigenvalue $G_{d,j}(\mathbf{k},\omega)$ and $\sigma$
is the internal degree of freedom. Explicitly, Eq.\,\eqref{s_3} becomes
\begin{eqnarray}
\text{Ch} &=& \frac{\varepsilon^{\mu\nu\rho}}{4\pi^{2}}\sum_{j,\sigma}\oint dz\int d^{2}ke^{z0^{+}}(G_{d,j}^{-1}\partial_{\mu}G_{d,j})(\partial_{\nu}[U^{-1}]_{j\sigma})(\partial_{\rho}U_{\sigma j}).\label{s_4}
\end{eqnarray}
\end{widetext}

If $G_{d,j}(\mathbf{k},\omega)$ has no zeros, but has poles below the
Fermi energy (i.e. the zero point), then around the pole $G_{d,j}(\mathbf{k},\omega)\sim\frac{\lambda(\mathbf{k},\omega)}{\omega-\varepsilon(\mathbf{k})}$,
we have 
\newpage
\begin{eqnarray}
&& \sum_{\sigma}(G_{d,j}^{-1}\partial_{\omega}G_{d,j})(\partial_{k_{x}}[U^{-1}]_{j\sigma})(\partial_{k_{y}}U_{\sigma j}) \notag\\
& & \sim \sum_{\sigma} \frac{-1}{\omega-\varepsilon(\mathbf{k})}[\partial_{k_{x}}\langle\psi_{j}(\mathbf{k},\omega)|\mathbf{k},\sigma\rangle][\partial_{k_{y}}\langle\mathbf{k},\sigma|\psi_{j}(\mathbf{k},\omega)\rangle] \nonumber \\
 & & =  \frac{-1}{\omega-\varepsilon(\mathbf{k})}\langle\partial_{k_{x}}\psi_{j}(\mathbf{k},\omega)|\partial_{k_{y}}\psi_{j}(\mathbf{k},\omega)\rangle\label{s_5},
\end{eqnarray}
and 
\begin{eqnarray}
&& \sum_{\sigma}(G_{d,j}^{-1}\partial_{k_{x}}G_{d,j})(\partial_{k_{y}}[U^{-1}]_{j\sigma})(\partial_{\omega}U_{\sigma j}) \notag \\
 & & \quad\quad  \sim \frac{\partial_{k_{x}}\varepsilon(\mathbf{k})}{\omega-\varepsilon(\mathbf{k})}\langle\partial_{k_{y}}\psi_{j}(\mathbf{k},\omega)|\partial_{\omega}\psi_{j}(\mathbf{k},\omega)\rangle. \label{s_6}
\end{eqnarray}
Thus, for each pole, using the residue theorem for Eq.\,\eqref{s_4}, we obtain 
\begin{eqnarray}
 &  & \frac{1}{2\pi i}\int d^{2}k\{\langle\partial_{k_{x}}\psi_{j}(\mathbf{k},\omega)|\partial_{k_{y}}\psi_{j}(\mathbf{k},\omega)\rangle \nonumber \\
 &  & \quad\quad\quad\quad\quad
+\partial_{k_{y}}\varepsilon(\mathbf{k})\langle\partial_{k_{x}}\psi_{j}(\mathbf{k},\omega)|\partial_{\omega}\psi_{j}(\mathbf{k},\omega)\rangle\nonumber \\
 &  & \quad\quad\quad\quad\quad+\partial_{k_{x}}\varepsilon(\mathbf{k})\langle\partial_{\omega}\psi_{j}(\mathbf{k},\omega)|\partial_{k_{y}}\psi_{j}(\mathbf{k},\omega)\rangle\}|_{\omega=\varepsilon(\mathbf{k})} \nonumber \\
 &  & \quad\quad\quad\quad\quad
-(k_{x}\leftrightarrow k_{y}).\label{s_7}
\end{eqnarray}
Defining the quasiparticle band $|\psi_{j}(\mathbf{k})\rangle=|\psi_{j}(\mathbf{k},\varepsilon(\mathbf{k}))\rangle$,
we can simplify Eq.\,{\eqref{s_7}} to  
\begin{equation}
\frac{1}{2\pi i}\int d^{2}k\{\langle\partial_{k_{x}}\psi_{j}(\mathbf{k})|\partial_{k_{y}}\psi_{j}(\mathbf{k})\rangle-(k_{x}\leftrightarrow k_{y}).
\end{equation}
Summing all of these quasiparticle bands below Fermi energy, we get the result  (eq.\eqref{eq:berry_ful}) given in the main text. 

If $G$ has zeros [note that Eq.\,\eqref{s_4} is dual for $G$ and $G^{-1}$], then around the zeros, $G_{d,j}(\mathbf{k},\omega)\sim\lambda(\mathbf{k},\omega)[\omega-\varepsilon(\mathbf{k})]$, a similar discussion can be used, and we obtain 
\begin{eqnarray}
&& \sum_{\sigma}(G_{d,j}^{-1}\partial_{\omega}G_{d,j})(\partial_{k_{x}}[U^{-1}]_{j\sigma})(\partial_{k_{y}}U_{\sigma j}) \notag \\
&& \quad\quad  \sim  \frac{1}{\omega-\varepsilon(\mathbf{k})}\langle\partial_{k_{x}}\psi_{j}(\mathbf{k},\omega)|\partial_{k_{y}}\psi_{j}(\mathbf{k},\omega)\rangle\label{s_8},
\end{eqnarray}
and 
\begin{eqnarray}
&& \sum_{\sigma}(G_{d,j}^{-1}\partial_{k_{x}}G_{d,j})(\partial_{k_{y}}[U^{-1}]_{j\sigma})(\partial_{\omega}U_{\sigma j}) 
 \notag \\
&& \quad\quad  \sim  -\frac{\partial_{k_{x}}\varepsilon(\mathbf{k})}{\omega-\varepsilon(\mathbf{k})}\langle\partial_{k_{y}}\psi_{j}(\mathbf{k},\omega)|\partial_{\omega}\psi_{j}(\mathbf{k},\omega)\rangle.\label{s_9}
\end{eqnarray}
The only difference from Eqs.\,\eqref{s_5} and \eqref{s_6} is the sign, which explains the result [eq.\eqref{berry_a}]  in the main text.

\section{$U(n)$ symmetry and Gram-Schmidt orthonormalization }

For the noninteracting case, $|\psi_{1}(\mathbf{k})\rangle, |\psi_{2}(\mathbf{k})\rangle, \cdots, $ $ |\psi_{n}(\mathbf{k})\rangle$ are orthogonal with each other. In the following, we prove that in this case, $\text{Ch}{}_{F}$ [eq.\,\eqref{eq:berry_ful} in the main text] is invariant under a smooth $U(n)$ transformation
$|\psi_{j}(\mathbf{k})\rangle=\sum_{l}U_{jl}(\mathbf{k})|\tilde{\psi}_{l}(\mathbf{k})\rangle$
[note that $U_{jl}(\mathbf{k})$ can be chosen to be well defined in the whole Brillouin zone].   Substituting
the transformation into the Berry curvature, we obtain 
\begin{eqnarray}
&& \sum_{j=1}^{n} \int d^{2}k\langle\partial_{k_{x}}\psi_{j}(\mathbf{k})|\partial_{k_{y}}\psi_{j}(\mathbf{k})\rangle \notag \\
&&   =   \sum_{j,l,l'=1}^{n}\int d^{2}k\langle\partial_{k_{x}}[U_{jl}\tilde{\psi}_{l}(\mathbf{k})]|\partial_{k_{y}}[U_{jl'}\tilde{\psi}_{l'}(\mathbf{k})]\rangle\nonumber \\
 & &  = \sum_{j,l,l'=1}^{n}\int d^{2}k\langle\partial_{k_{x}}\tilde{\psi}_{l}(\mathbf{k})|\partial_{k_{y}}\tilde{\psi}_{l'}(\mathbf{k})\rangle U_{jl}^{*}U_{jl'}\nonumber \\
 &  & \quad  +\sum_{j,l,l'=1}^{n}\int d^{2}k\langle\tilde{\psi}_{l}(\mathbf{k})|\partial_{k_{y}}\tilde{\psi}_{l'}(\mathbf{k})\rangle(\partial_{k_{x}}U_{jl}^{*})U_{jl'}\nonumber \\
 &  &\quad  +\sum_{j,l,l'=1}^{n}\int d^{2}k\langle\partial_{k_{x}}\tilde{\psi}_{l}(\mathbf{k})|\tilde{\psi}_{l'}(\mathbf{k})\rangle(U_{jl}^{*}\partial_{k_{y}}U_{jl'})\nonumber \\
 &  &\quad  +\sum_{j,l,l'=1}^{n}\int d^{2}k\langle\tilde{\psi}_{l}(\mathbf{k})|\tilde{\psi}_{l'}(\mathbf{k})\rangle\partial_{k_{x}}U_{jl}^{*}\partial_{k_{y}}U_{jl'}. \label{eq4}
\end{eqnarray}
Using the unitary of $U$, we get
\begin{eqnarray}
&& \sum_{j,l,l'=1}^{n}\int d^{2}k\langle\partial_{k_{x}}\tilde{\psi}_{l}(\mathbf{k})|\partial_{k_{y}}\tilde{\psi}_{l'}(\mathbf{k})\rangle U_{jl}^{*}U_{jl'}\notag \\
&&  \quad\quad =\sum_{l}^{n}\int d^{2}k\langle\partial_{k_{x}}\tilde{\psi}_{l}(\mathbf{k})|\partial_{k_{y}}\tilde{\psi}_{l}(\mathbf{k})\rangle, 
\end{eqnarray}
and
\begin{eqnarray}
&& \sum_{j,l,l'=1}^{n}\int d^{2}k\langle\tilde{\psi}_{l}(\mathbf{k})|\partial_{k_{y}}\tilde{\psi}_{l'}(\mathbf{k})\rangle(\partial_{k_{x}}U_{jl}^{*})U_{jl'} \notag \\
&&   =\sum_{j,l,l'=1}^{n}\int d^{2}k\langle\partial_{k_{y}}\tilde{\psi}_{l}(\mathbf{k})|\tilde{\psi}_{l'}(\mathbf{k})\rangle(U_{jl}^{*}\partial_{k_{x}}U_{jl'}).  \label{s_12}
\end{eqnarray}
Combining Eq.\, \eqref{s_12} with the third term of Eq.\,\eqref{eq4}, and using the antisymmetry for the exchange of $k_{x}$ and $k_{y}$, we find that the second and third terms of Eq.\,\eqref{eq4} will not contribute to $\text{Ch}_{F}$. The last term in  Eq.\,\eqref{eq4} is
\begin{eqnarray}
&& \sum_{j,l,l'=1}^{n}\int d^{2}k\langle\tilde{\psi}_{l}(\mathbf{k})|\tilde{\psi}_{l'}(\mathbf{k})\rangle\partial_{k_{x}}U_{jl}^{*}\partial_{k_{y}}U_{jl'} \notag \\
  = &&\sum_{j,l=1}^{n}\int d^{2}k\partial_{k_{x}}U_{jl}^{*}\partial_{k_{y}}U_{jl} \notag\\
 =  &&   \sum_{j,l=1}^{n}\int d^{2}k\partial_{k_{x}}(U_{jl}^{*}\partial_{k_{y}}U_{jl})   -\sum_{j,l=1}^{n}\int d^{2}kU_{jl}^{*}\partial_{k_{x}}\partial_{k_{y}}U_{jl}, \notag 
\end{eqnarray}
the first term of which vanishes due to the periodic boundary condition and the second term of which will be canceled by the antisymmetry in $k_{x},k_{y}$. Finally, we have
\begin{eqnarray}
&& \sum_{j=1}^{n}\int d^{2}k\langle\partial_{k_{x}}\psi_{j}(\mathbf{k})|\partial_{k_{y}}\psi_{j}(\mathbf{k})\rangle \notag \\
&&  \quad\quad =\sum_{l}^{n}\int d^{2}k\langle\partial_{k_{x}}\tilde{\psi}_{l}(\mathbf{k})|\partial_{k_{y}}\tilde{\psi}_{l}(\mathbf{k})\rangle,
\end{eqnarray}
showing the $U(n)$ symmetry.

Before going to the interacting case, we want to stress that
\begin{equation}
\int d^{2}k[\langle\partial_{k_{x}}\psi_{j}(\mathbf{k})|\partial_{k_{y}}\psi_{j}(\mathbf{k})\rangle-\langle\partial_{k_{y}}\psi_{j}(\mathbf{k})|\partial_{k_{x}}\psi_{j}(\mathbf{k})\rangle]
\end{equation}
is quantized and it is stable for a smooth deformation of $|\psi_{j}(\mathbf{k})\rangle$ \cite{Bernevig2013book}.

For the interacting case, $|\psi_{1}(\mathbf{k})\rangle,|\psi_{2}(\mathbf{k})\rangle,\cdots,|\psi_{n}(\mathbf{k})\rangle$
are not orthogonal to each other. In the following, we prove  that these states can be smoothly deformed to an orthogonal basis without changing $\text{Ch}_F$, supposing that they are still linear independent. Using Gram-Schmidt orthogonalization, we define an orthogonal orthogonal basis:  $|\Psi_{1}(\mathbf{k})\rangle=|\psi_{1}(\mathbf{k})\rangle$, $|\Psi_{2}(\mathbf{k})\rangle=\frac{|\psi_{2}(\mathbf{k})\rangle-\langle\Psi_{1}(\mathbf{k})|\psi_{2}(\mathbf{k})\rangle|\Psi_{1}(\mathbf{k})\rangle}{|||\psi_{2}(\mathbf{k})\rangle-\langle\Psi_{1}(\mathbf{k})|\psi_{2}(\mathbf{k})\rangle|\Psi_{1}(\mathbf{k})\rangle||}$, $\cdots$. It is easy to check that  $\langle\Psi_{j}(\mathbf{k})|\psi_{j}(\mathbf{k})\rangle = \langle\psi_{j}(\mathbf{k})|\Psi_{j}(\mathbf{k})\rangle > 0$ for $j=1,\cdots,n$. The smooth deformation then is defined by ($t\in[0,1]$)
\begin{equation}
|\Psi_{i}(\mathbf{k},t)\rangle=\frac{(1-t)|\psi_{i}(\mathbf{k})\rangle+t|\Psi_{i}(\mathbf{k})\rangle}{||(1-t)|\psi_{i}(\mathbf{k})\rangle+t|\Psi_{i}(\mathbf{k})\rangle||}.\label{eq6}
\end{equation}
This deformation is well defined since $
||(1-t)|\psi_{i}(\mathbf{k})\rangle+t|\Psi_{i}(\mathbf{k})\rangle|| = (1-t)^{2}+t^{2}+(1-t)t[\langle\psi_{i}(\mathbf{k})|\Psi_{i}(\mathbf{k})\rangle+\langle\Psi_{i}(\mathbf{k})|\psi_{i}(\mathbf{k})\rangle] >0$.
Thus, we have defined a smooth deformation, so that $|\psi_{1}(\mathbf{k})\rangle$, $|\psi_{2}(\mathbf{k})\rangle$, $\cdots$, $|\psi_{n}(\mathbf{k})\rangle$
go to $|\Psi_{1}(\mathbf{k})\rangle$, $|\Psi_{2}(\mathbf{k})\rangle$, $\cdots$, $|\Psi_{n}(\mathbf{k})\rangle$ without changing the Chern number. And using the $U(n)$ symmetry of the Chern number for the orthogonal basis, we get the conclusion that the Chern number is  related only to the space spanned by $|\psi_{1}(\mathbf{k})\rangle,|\psi_{2}(\mathbf{k})\rangle,\cdots,|\psi_{n}(\mathbf{k})\rangle$.

\section{Time-reversal-invariant system}

For a $\mathcal{T}$-symmetric system,  it is easy to check that if $|\psi_{n}\rangle$ is a set of complete orthogonal bases, then $\mathcal{T}|\psi_{n}\rangle$ is too. On the
other hand, for any $|\psi\rangle$ and $|\phi\rangle$, we have $\langle\psi|\mathcal{T}^{-1}|\phi\rangle=(\langle\mathcal{T}\psi|\phi\rangle)^{*}$. Using the definitions $c_{i\sigma}(\tau)=e^{\tau{H}}c_{i\sigma}e^{-\tau {H}}$, $c_{i\sigma}(t)=e^{i{H}t}c_{i\sigma}e^{-i{H}t}$, and $\mathcal{T}c_{i}\mathcal{T}^{-1}=i\sigma_{y}c_{i}$, where $H$ is the many-body Hamiltonian, $c_{i}=(c_{i\uparrow},c_{i\downarrow})^{T}$, and the index $i$ includes the lattice index and other internal degrees of freedom besides spin, we have 
\begin{eqnarray}
G_{\downarrow,ij}(\tau) &=& -\sum_{n}\langle\psi_{n}|T_{\tau}[c_{i\downarrow}(\tau)c_{j\downarrow}^{\dag}(0){e}^{-\beta ({H}-\Omega)}]|\psi_{n}\rangle \notag\\ 
&=& - \sum_{n}\langle\psi_{n}|\mathcal{T}^{-1}T_{\tau}[c_{i\uparrow}(\tau)c_{j\uparrow}^{\dag}(0){e}^{-\beta ({H}-\Omega)}]\mathcal{T}|\psi_{n}\rangle \notag\\
&=&- (\sum_{n}\langle\mathcal{T}\psi_{n}|T_{\tau}[c_{i\uparrow}(\tau)c_{j\uparrow}^{\dag}(0){e}^{-\beta ({H}-\Omega)}]|\mathcal{T}\psi_{n}\rangle)^{*}  \notag\\
&=& G_{\uparrow,ij}^{*}(\tau).
\end{eqnarray}
Here, $e^{-\beta \Omega}=\sum_{n}\langle\psi_{n}|{e}^{-\beta {H}}|\psi_{n}\rangle$. Similarly, for the retarded Green's function, 
\begin{eqnarray}
G_{\text{ret},ij}^{\uparrow}(t)&=&-i\Theta(t)\langle\psi_{n}|\{c_{i\uparrow}(t),c_{j\uparrow}^{\dag}(0)\} {e}^{-\beta ({H}-\Omega)}|\psi_{n}\rangle \notag\\
&=&-i\Theta(t)\langle\psi_{n}|\mathcal{T}^{-1}\{c_{i\downarrow}(-t),c_{j\downarrow}^{\dag}(0)\}{e}^{-\beta ({H}-\Omega)} \mathcal{T} |\psi_{n}\rangle \notag \\
&=&-i\Theta(t)(\langle\psi_{n}| \{c_{i\downarrow}(-t),c_{j\downarrow}^{\dag}(0)\} {e}^{-\beta ({H}-\Omega)}|\psi_{n}\rangle)^{*} \notag\\
&=&-i\Theta(t)\langle\psi_{n}|  \{c_{j\downarrow}(0),c_{i\downarrow}^{\dag}(-t)\} {e}^{-\beta ({H}-\Omega)}|\psi_{n}\rangle \notag \\
&=& G_{\text{ret},ji}^{\downarrow}(t).
\end{eqnarray}
After that, using Fourier transformation for position and time, we get 
$G_{\uparrow}^{*}(\mathbf{k}, i\omega)=G_{\downarrow}(-\mathbf{k},-i\omega)$,
$\rho_\uparrow^*(\mathbf{k}) = \rho_\downarrow(-\mathbf{k})$, and $G_{\uparrow}(\mathbf{k}, \omega+i\delta)=G^T_{\downarrow}(-\mathbf{k}, \omega+i\delta)$.

\begin{figure}
\includegraphics[width=3in]{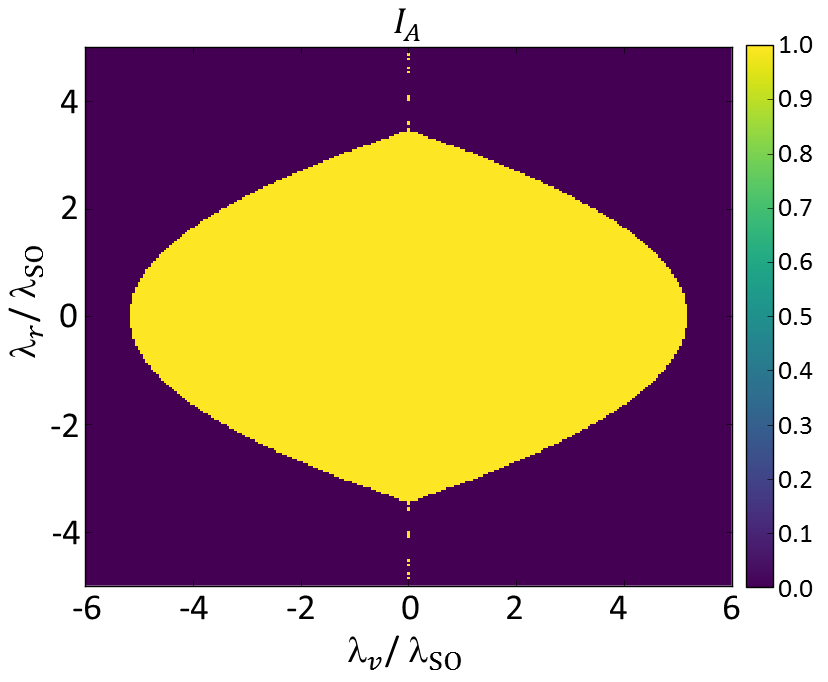} 
\caption{The topological number $\text{I}_{A}$ for the spin-up subsystem}\label{subsystem}
\end{figure}

The numerical results for the topological
index ${I}_{A}$ of the spin-up subsystem of the Kane-Mele model is shown in Fig.\,\ref{subsystem}.

\section{The density-matrix description for the interacting case}
We treat $G_{A}^{\sigma\sigma^{\prime}}(\mathbf{k},\tau-\tau^{\prime})=-\langle T_{\tau}\hat{c}_{A,\mathbf{k}\sigma}(\tau)\hat{c}_{A,\mathbf{k}\sigma^{\prime}}^{\dagger}(\tau^{\prime})\rangle$ in the full system. The Lehmann representation of the
Green's function in the zero-temperature limit is \cite{Wang2012prx}
\begin{eqnarray}
G_{A}^{\sigma\sigma^{\prime}}(\mathbf{k},\omega) &=& \sum_{\tilde{j}}\frac{\langle0|c_{A,\mathbf{k}\sigma}|\tilde{j}\rangle\langle \tilde{j}|c_{A,\mathbf{k}\sigma'}^{\dagger}|0\rangle}{\omega-(E_{ \tilde{j}}-E_{0})} \notag \\
&&+\sum_{{j}}\frac{\langle {j}|c_{A,\mathbf{k}\sigma}|0\rangle\langle0|c_{A,\mathbf{k}\sigma'}^{\dagger}|{j}\rangle}{\omega+(E_{{j}}-E_{0})},\label{green}
\end{eqnarray}
where $|j\rangle$ and $|\tilde j\rangle$ are the many-body
wave functions with energies $E_j$ and  $E_{\tilde j}$ and $|0\rangle$ is the ground state of the full system with energy $E_0$. For  an $M$ particle system,  $|j\rangle$ refers to the $(M-1)$-particle state, and  $|\tilde j\rangle$ refers to the $(M+1)$-particle state. Note that $-E_j + E_0 <0$ and $E_{\tilde j} -E_0 > 0$.  By defining vectors in the $A$ subspace $|\varphi_j \rangle$ and $|\varphi_{\tilde{j}} \rangle$ with $\langle \mathbf{k}, \sigma |\varphi_j \rangle \equiv  \langle j |c_{A,\mathbf{k}\sigma}|{0}\rangle $ and   $\langle \mathbf{k}, \sigma |\varphi_{\tilde{j}} \rangle \equiv  \langle0|c_{A,\mathbf{k}\sigma}|\tilde{j}\rangle $, we have 
\begin{equation}
G_{A}(\mathbf{k},\omega)=\sum_{j}\frac{|\varphi_{{j}} \rangle  \langle\varphi_{{j}} | }{\omega-(-E_{j}+E_{0})}+\sum_{\tilde{j}}\frac{|\varphi_{\tilde{j}} \rangle  \langle\varphi_{\tilde{j}} |}{\omega-(E_{\tilde{j}}-E_{0})}.\label{green1}
\end{equation}
Without interaction and coupling, the union of $\{\varphi_j\}$ and $\{\varphi_{\tilde{j}}\}$ is an orthogonal set. For the case with interactions or coupling with environments, they are not orthogonal to each other any longer.  For the weak-coupling and weak-interaction case, we can move all of  $-E_j + E_0$ together to $\varepsilon_G < 0$ and all of $E_{\tilde j} - E_0$ to  $\varepsilon_E > 0$ without the bands and blind bands crossing the zero point, and finally we get 
\begin{equation}
\underline{G}_{A}(\mathbf{k},\omega)=\frac{\rho_A(\mathbf{k}) }{\omega-\varepsilon_G}+\frac{\bar{\rho}_A(\mathbf{k})}{\omega-\varepsilon_E},\label{green2}
\end{equation}
where $\rho_A(\mathbf{k})=\sum_{j} |\varphi_{{j}} \rangle  \langle\varphi_{{j}} |$ and $\bar{\rho}_A(\mathbf{k}) = \sum_{\tilde{j}} |\varphi_{\tilde{j}} \rangle  \langle\varphi_{\tilde{j}}|$. In the following, we show that $\bar{\rho}_A(\mathbf{k}) = \textbf{I}_A - {\rho}_A(\mathbf{k})$, so that all of the discussion about the noninteracting case can be applied to the weak interacting system.

By defining vectors in the full space ($A+E$), $|\psi_j \rangle $ and $|\psi_{\tilde{j}} \rangle $, with $\langle \mathbf{k}, \sigma |\psi_j \rangle \equiv  \langle j|c_{A,\mathbf{k}\sigma}|{0}\rangle $,  $\langle \mathbf{k}, \eta |\psi_j \rangle \equiv  \langle j |c_{E,\mathbf{k}\eta}|{0}\rangle $,   $\langle \mathbf{k}, \sigma |\psi_{\tilde{j}} \rangle \equiv  \langle 0 |c_{A,\mathbf{k}\sigma}|\tilde{j}\rangle $, and  $\langle \mathbf{k}, \eta |\psi_{\tilde{j}} \rangle \equiv  \langle0|c_{E,\mathbf{k}\eta}|\tilde{j}\rangle $, we have  $ |\varphi_j \rangle= P_A|\psi_j \rangle$, $ |\varphi_{\tilde{j}} \rangle= P_A|\psi_{\tilde{j}} \rangle$. The Green's function for the full system becomes:
\begin{eqnarray}
G_{F}(\mathbf{k},\omega+i\delta) &=& \sum_{j}\frac{|\psi_{{j}} \rangle  \langle\psi_{{j}} | }{\omega+i\delta-(-E_{j}+E_{0})} \notag \\
&& +\sum_{\tilde{j}}\frac{|\psi_{\tilde{j}} \rangle  \langle\psi_{\tilde{j}} |}{\omega+i\delta-(E_{\tilde{j}}-E_{0})}.\label{green3}
\end{eqnarray}
For any vector $|\mathbf{k}, l\rangle$ in the full space ($A+E$),  the spectral function of $G_F$ satisfies \cite{Mahan}
\begin{equation}
1= -\int_{-\infty}^{\infty} d\omega\frac{1}{\pi}\text{Im} [\langle \mathbf{k}, l | G_{F}(\mathbf{k},\omega+i\delta)|\mathbf{k}, l\rangle],
\label{green4}
\end{equation}
which implies 
\begin{equation}
\mathbf{I}=\sum_{j}{|\psi_{{j}} \rangle  \langle\psi_{{j}} | }+\sum_{\tilde{j}}{|\psi_{\tilde{j}} \rangle  \langle\psi_{\tilde{j}} |}.\label{green6}
\end{equation}
By projecting on the $A$ subspace, we get the result $\bar{\rho}_A(\mathbf{k}) = \textbf{I}_A - {\rho}_A(\mathbf{k})$.

\end{document}